\documentclass[journal,twoside]{./IEEEtran}

\usepackage[pdftex]{graphicx}
\DeclareGraphicsExtensions{.pdf}
\usepackage{xspace}
\usepackage{setspace} 

\usepackage[cmex10]{amsmath}
\usepackage{amsfonts}
\usepackage{amssymb}
\usepackage{algpseudocode}

\usepackage{xcolor}
\usepackage{cite}
\usepackage{listings}
\usepackage{chngcntr}
\counterwithout{paragraph}{subsubsection}
\counterwithin*{paragraph}{section}

\newcommand{\figref}[1]{{Fig.~\ref{fig:#1}}}
\newcommand{\seclabel}[1]{\label{sec:#1}}
\newcommand{\figlabel}[1]{\label{fig:#1}}
\newcommand{\secref}[1]{\mbox{Section{ }\ref{sec:#1}}}

\newcommand{\noop}[1]{}

\def\1{{\mathbf 1}}        

\renewcommand{\baselinestretch}{0.91} 

\title{Platform-Based Design Methodology and Modeling for Aircraft Electric Power Systems}

\author{\IEEEauthorblockN{
Pierluigi Nuzzo$^*$, John Finn$^*$, Mohammad Mozumdar$^{\S}$ and 
Alberto Sangiovanni-Vincentelli$^*$}\\
\IEEEauthorblockA{* Department of Electrical Engineering and Computer Science, 
University of California at Berkeley \\
$^{\S}$ Department of Electrical Engineering, California State University Long Beach}
}

\begin{document}

\markboth{Proceedings of Green Energy and Systems Conference 2013, November 25, Long Beach, CA, USA. }{This full text paper was peer reviewed at the direction of Green Energy and Systems Conference subject matter experts.}

\maketitle

\begin{abstract}
In an aircraft electric power system (EPS), a supervisory control unit must
actuate a set of switches to distribute power from generators to loads, while
satisfying safety, reliability and real-time performance requirements.
To reduce expensive re-design steps in current design methodologies, such a
control problem is generally addressed based on minor incremental changes on top
of consolidated solutions, since it is difficult to estimate the impact of
earlier design decisions on the final implementation. In this paper, we
introduce a methodology for the design space exploration and virtual prototyping
of EPS supervisory control protocols, following the platform-based design (PBD)
paradigm. Moreover, we describe the modeling infrastructure that supports the
methodology.
In PBD, design space exploration is carried out as a sequence of refinement
steps from the initial specification towards a final implementation, by mapping
higher-level behavioral models into a set of library components at a lower level
of abstraction. In our flow, the system specification is captured using SysML
requirement and structure diagrams. State-machine diagrams enable verification
of the control protocol at a high level of abstraction, while lower-level hybrid
models, implemented in Simulink, are used to verify properties related to
physical quantities, such as time, voltage and current values. The effectiveness
of our approach is illustrated on a prototype EPS control protocol design.
\end{abstract}

\section{Introduction}
\thispagestyle{empty}

The advent of high capability, reliable power electronics together with powerful
embedded processors has enabled, over the last fifteen years, an increasing
amount of ``electrification'' of vehicles such as cars and
aircraft~\cite{Moir08,Pinto2010}. In an aircraft, hydraulic, pneumatic and
mechanical systems are replaced by cyber-electrical components improving the
overall system efficiency. However, the increase of electrically-powered
elements poses significant challenges to the aircraft electric power system
(EPS) in terms of power generation and distribution under tight reliability and
safety constraints for the cyber-electric components.

A severe limitation in common design practices for such kind of systems 
is the lack of formal
specifications. Requirements are often written in languages that are not
suitable for mathematical analysis and verification.
Assessing system correctness is then left for simulations later in the design
process and prototype tests. The inability to rigorously model the interactions
among heterogeneous components and between the ``physical'' and the ``cyber''
aspects of the system is also a serious obstacle. Thus, the traditional
heuristic design process based on informal requirement capture and designers'
experience leads to implementations that are inefficient and sometimes do not
even satisfy the requirements yielding long re-design cycles, cost overruns and
unacceptable delays.


We propose instead to carry out the design with a rigorous
flow that selects available components using an
optimization process including allocation of
requirements to the components and early validation of
design constraints made possible by the formalization we
advocate. Our methodology follows the platform-based design
(PBD) paradigm~\cite{ASV:quovadis}, which
has been successfully adopted in the automotive and consumer
electronics~\cite{nuzzo:kluPipeline} domains to overcome
similar challenges, by formalizing the design flow as a
sequence of refinement steps from the original specification
to the final implementation.

A basic principle of PBD is the distinction between the \emph{function} (what
the system is supposed to do, i.e.~the specifications) and the
\emph{architecture} (how specifications are realized, i.e.~the components
implementing the function together with their interconnections) at each
abstraction level, which allows for automatic design space exploration. At each
refinement step, the design is regarded as a platform instance, i.e.~a valid
composition of library elements that are pre-characterized by their cost and
performance metrics. The objective is therefore to select a platform instance
that correctly implements a given specification. The mapping of such a
specification onto an architecture can be formalized by an optimization problem
whose solution represents the functional specification to be implemented by the
subsequent refinement step. This process repeats recursively until an
implementation is reached.

A key element for the successful deployment of PBD is the definition of a set of
appropriate abstraction layers for efficient and accurate system exploration as
well as the generation of a rich set of models, which represent different
viewpoints (aspects) of the design, and can be used by different,
domain-specific analysis and verification tools. In this paper, we present a
design methodology based on PBD and a supporting modeling infrastructure to be
used for requirement capture, simulation and virtual prototyping of EPS
supervisory control protocols. We explore the capabilities of the Systems
Modeling Language\footnote{SysML is an object oriented modeling language largely
based on the Unified Modeling Language (UML) 2.1, which also provides useful
extensions for systems engineering.}~\cite{sysml} (SysML) to represent the EPS
\emph{function}, and the Simulink SimPowerSystem library to assemble the EPS
\emph{architecture}. System specifications are first categorized in terms of
safety, reliability and performance constraints. SysML requirement and
state-machine diagrams capture such specifications and allow verification of the
design at a high level of abstraction. On the other hand, hybrid models
implemented in Simulink allow expressing and verifying a set of properties at a
lower abstraction level, including timing and predicates on current and voltage
signals.


Our methodology builds on a number of results that have opened the way for a
more structured approach to EPS design. The adoption of model-based development
(MBD) and simulation for the analysis of aircraft performance has already been
advocated in~\cite{Krus00,Bals05}. In the context of the More Open Electrical
Technologies (MOET) project~\cite{MOET09}, a set of model libraries have been
developed using the Modelica language~\cite{modelica} to support
``more-electric'' aircraft simulation, design and validation. Simulation is used
for electric power system performance verification (e.g., stability and power
quality) at the network level, by leveraging models with different levels of
complexity to analyze different system properties, and validated with real
equipment measurements.
However, design space exploration, optimization and analysis of faulty behaviors
in these models can still become computationally unaffordable unless proper
levels of abstraction are devised, based on the goals at each design step.

Our methodology has similarities with~\cite{Pinto2010}, which deals with how to
select the power generators and synthesize the EPS topology (interconnection
among the various components) by formulating and solving a set of binary
optimization problems.
However, not all the requirements of an EPS can be accurately approximated by
binary or mixed integer-linear constraints, which calls for an extension of the
flow in~\cite{Pinto2010}, to enable synthesis of EPS control protocols subject
to heterogeneous sets of system constraints. A number of recent papers have
investigated the adoption of SysML for MBD and analysis of complex systems of
systems, albeit not within a rigorously formalized methodology.
In~\cite{huynh2006} and~\cite{Guo2009} the adoption of SysML is investigated for
MBD of maritime and automotive systems, while in~\cite{Johnson2007} a language
extension is proposed to support the continuous-time (CT) dynamics of physical
systems using the Modelica language.
While using Simulink CT models to enrich our SysML library, our approach is
complementary to the one in~\cite{Johnson2007}, since our main focus is on the
methodology rather than on the modeling languages themselves.
%


We first describe the electric power system that is considered in this paper in
\secref{eps}. \secref{flow} details our design methodology and model library.
\secref{application} reports results from the application of our methodology to
a prototype design. Concluding remarks follow in \secref{conclusions}.

\section{The Aircraft Electric Power System}
\seclabel{eps}

\begin{figure}[t]
\centering
\includegraphics[width=0.9\columnwidth]{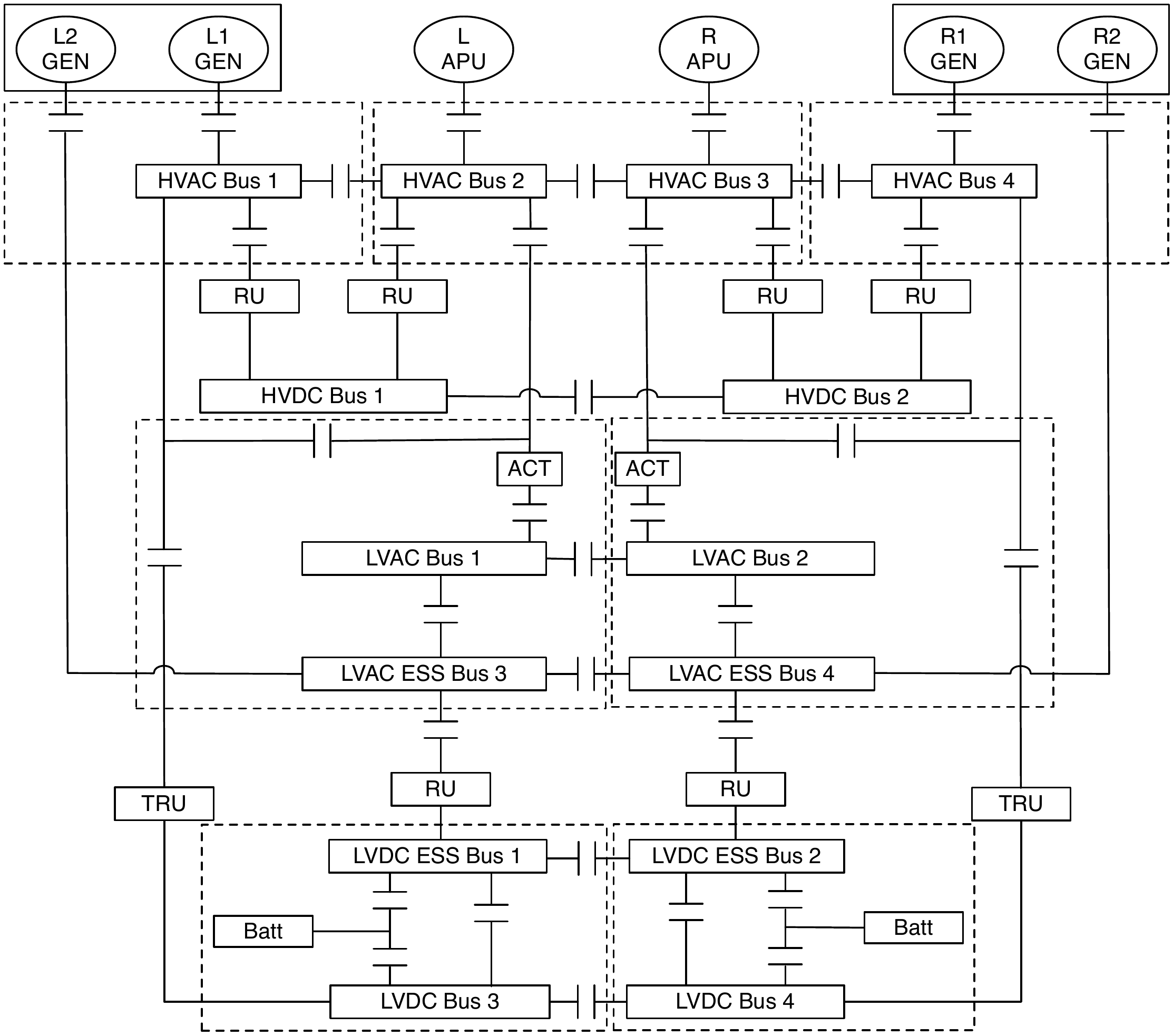}
\caption{Single-line diagram of an aircraft EPS adapted from
a Honeywell, Inc. patent~\cite{Michalko2008}.}
\label{fig:typEPS}
\end{figure}

Fig.~\ref{fig:typEPS} illustrates a sample architecture for
power generation and distribution in a passenger aircraft in the form of
a single-line diagram~\cite{Moir08}, a simplified notation 
for three-phase power systems. 

\paragraph{Components}

The main components of an EPS schematic are generators, contactors, buses, and
loads. AC \emph{generators} supply power to buses, and can operate at either
high or low-voltages. Primary generators are connected to the aircraft engine,
while auxiliary generators, mounted on top of the \emph{Auxiliary Power Units}
(APU) or \emph{batteries} are used in flight when one of the primary generators
fails. AC and DC power \emph{buses} deliver power to a number of loads. Buses
can be essential or non-essential. Essential buses supply loads that cannot be
unpowered for more than a specified time interval, while non-essential buses
supply loads that may be shed in the case of a fault.
\emph{Contactors} are electromechanical switches that
establish connections between components, and therefore
determine the power flow from sources to buses and loads.
They are configured to be open or closed by one or multiple
supervisory controllers, generically denoted here as \emph{Bus Power Control Unit}
(BPCU).
\emph{Loads} include sub-systems such as lighting, heating, avionics, 
navigation as well as power conversion devices.
\emph{Rectifier units} (RU) convert AC power to DC power,
while AC \emph{transformers} (ACT) step down a high-voltage
to a lower one. Finally, combined \emph{Transformer Rectifier
Units} (TRU) both decrease the voltage level and convert it
from AC to DC.

\paragraph{System Description}

The main AC power sources at the top of the diagram include two low-voltage
generators, two high-voltage generators, and two APU generators. Each engine
connects to a high-voltage AC (HVAC) generator (L1 and R1) and a low-voltage AC
(LVAC) emergency generator (L2 and R2). Panels, denoted as dashed square boxes,
represent groups of components that are physically separated on the aircraft.
The three panels below the generators include the HVAC distribution buses, which
can be selectively connected to the HVAC generators, to the APUs, and to each
other via contactors, denoted by double bars.

The two panels below the high-voltage DC (HVDC) buses include the LVAC
sub-system of the EPS. A set of transformers convert HVAC power to LVAC power
and are connected to four LVAC buses.
LVAC ESS Bus 3 and LVAC ESS bus 4 are essential and are selectively connected to
the two emergency generators.
Moreover, the LVAC essential buses are also connected to the RUs, converting the
LVAC power to low-voltage DC (LVDC) power. There are four LVDC buses in total,
each with essential and non-essential loads, as well as two batteries, which may
be selectively connected. Power can also be routed directly from the HVAC bus to
the LVDC buses 3 and 4 using TRUs.

A BPCU (which is not shown in \figref{typEPS}) controls the state (open or
closed) of the contactors and reconfigures the system based on the status and
availability of the power sources. A Generator Control Unit (GCU), inside each
generator, regulates its output voltage level to be within a specified range.
Fluctuations in the power required by the loads can be directly handled by the
GCU within the generator's power rating. Whenever the power demand exceeds the
generator's capability, the BPCU is responsible for possibly shedding
non-essential loads or rerouting some of them to another power source.

\paragraph{System Requirements}

Given a set of loads, together with
their power and reliability requirements, the goal is to
determine the system's architecture and control strategy  such
that the demand of the loads is satisfied for all flight
conditions and a set of predetermined faults. To better
formalize this design objective, we begin with a qualitative
analysis of the main system requirements, by categorizing them
in terms of safety, reliability and performance
requirements. For each of these categories, we provide a few
examples, which serve as a reference for the rest of
the paper.

\emph{Safety} specifications constrain the way each bus can be powered to avoid
loss of essential features.
For instance, to avoid generator damage, we prescribe that AC sources should
never be paralleled, i.e.~no AC bus can be powered by multiple generators at the
same time. Moreover, we require that essential loads (such as flight-critical
actuators) and buses never be unpowered for more than a time interval $t_{max}$.
%
%

\emph{Reliability} specifications describe the bounds on the
failure probabilities that can be tolerated for different
portions of the system. Based on its failure modes, every
EPS component is characterized by a failure rate $\lambda$, 
indicating that a failure occurs, on average, every
$1/\lambda$ hours.
Based on the component failure rates, a typical specification would require that
the failure probability for any essential load (i.e. the probability of being
unpowered for more than $t_{max}$) be smaller than $10^{-9}$~\cite{Moir08,arp}. 
Therefore, both the
controller and the EPS topology should be designed to accommodate any possible
combination of faults, potentially causing the failure of an essential
component, and having a joint probability larger than $10^{-9}$. In practice,
the reliability of an EPS is directly linked to the amount of redundant
components and paths in its topology. The reliability specifications will then
determine the combination of simultaneous faults that need to be accounted for
by the control protocol.

\emph{Performance} requirements specify quality metrics that
are desired for the system. For instance, each
bus is assigned a priority list determining in which order
available generators should be selected to power it. If the
first generator in the list is unavailable, then the bus
will be powered by the second generator, and so on. A
hypothetical prioritization list for the HVAC Bus 1 would
require, for instance, that L1 GEN has the priority, if
available. Otherwise, Bus 1 should receive power from the R1
GEN, then from the L APU generator, and finally from the R APU generator. In a
similar way, load management policies are based on
priority tables requiring, for instance, that the available
power be first allocated to the non-sheddable loads and then
to the sheddable loads, in a prescribed order.

\section{The Structure of the Methodology}
\seclabel{flow}

\begin{figure}[t]
\includegraphics[width=0.9\columnwidth]{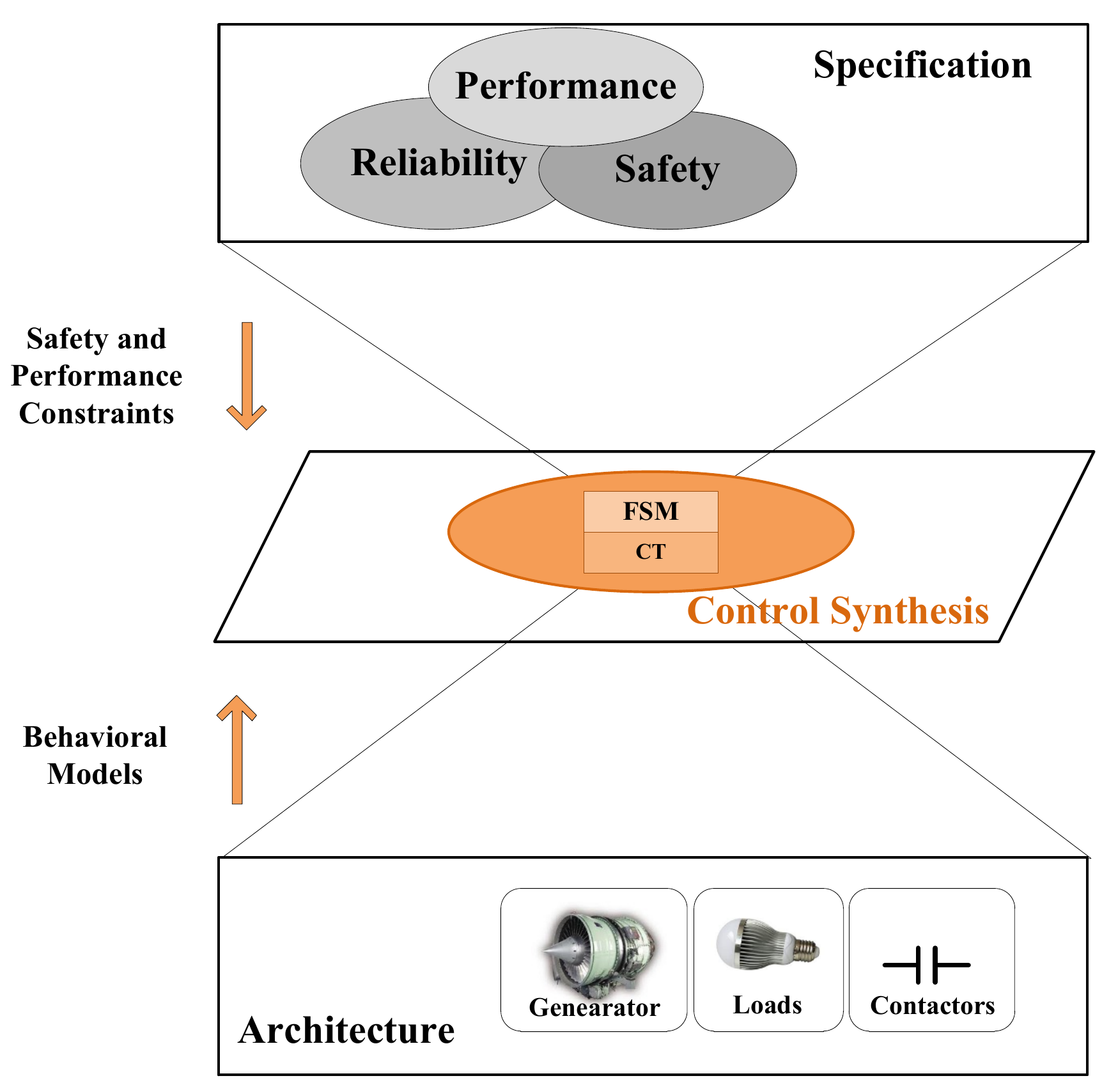}
\caption{EPS supervisory control design flow.}
\label{fig:flow}
\end{figure}

Given a set of requirements and a reference topology, the problem is to
design the BPCU state-machine to drive the contactors, while guaranteeing that
essential loads are correctly powered. 
Our methodology, represented in \figref{flow}, includes
a bottom-up and a top-down phase. In the bottom-up phase, we build a library of
platform elements, including the components described in \secref{eps} or some
aggregations of them. Each component is abstracted into behavioral and
performance models. Performance models characterize the physical attributes of a
component such as weight, size, power and cost. Behavioral models are organized
hierarchically to span different levels of abstractions, from finite-state
machine (FSM) abstract representations to continuous-time (CT) high-fidelity
models. In the top-down phase, we formalize system requirements in terms of
properties and constraints on the parameters and behaviors of the above models.
We formulate the EPS design exploration as an optimization problem where we
search the design space for candidate system configurations that satisfy the
conjunction of all the system constraints, and optimize some performance and
complexity (e.g.~number of components or states) metrics.

Based on the results in~\cite{Wongpiromsarn11,CDC_Mumu}, we assume that a
reactive control protocol can be synthesized from Linear Temporal Logic
(LTL)~\cite{Pnueli77} constructs and made available as a candidate BPCU state
machine for our exploration framework. Other approaches to control synthesis,
based on constrained optimization, can also be used to provide an initial 
candidate. Based on this assumption,
we can generate candidate BPCU designs and use our model library to verify the
design correctness with respect to the requirements. A set of specialized
analysis frameworks, denoted as \emph{theory managers} can be used to reason
about properties of different models, usually expressed using different formalisms,
following a similar paradigm as in satisfiability-modulo-theories
solvers~\cite{Hang11}. Each theory manager is responsible of executing a
specific system view and of verifying its correctness.  

Design space exploration is then organized as follows. 
%
The set of high-level EPS specifications and the topology are used to synthesize
an initial controller that satisfies safety and reliability requirements.
However, no notion of the physical constraints (e.g.,~timing, energy
consumption) related to the plant and the hardware implementation of the control
algorithm are available at this level of abstraction. Therefore, 
both the topology and the synthesized controller are executed using high-fidelity 
CT (or hybrid) models to assess the satisfaction of all requirements.
Simulation traces are monitored to both verify and optimize the controller. When all
requirements are satisfied, the candidate controller is returned as the final
design.

\begin{figure}[t]
\centering
\includegraphics[width=0.7\columnwidth]{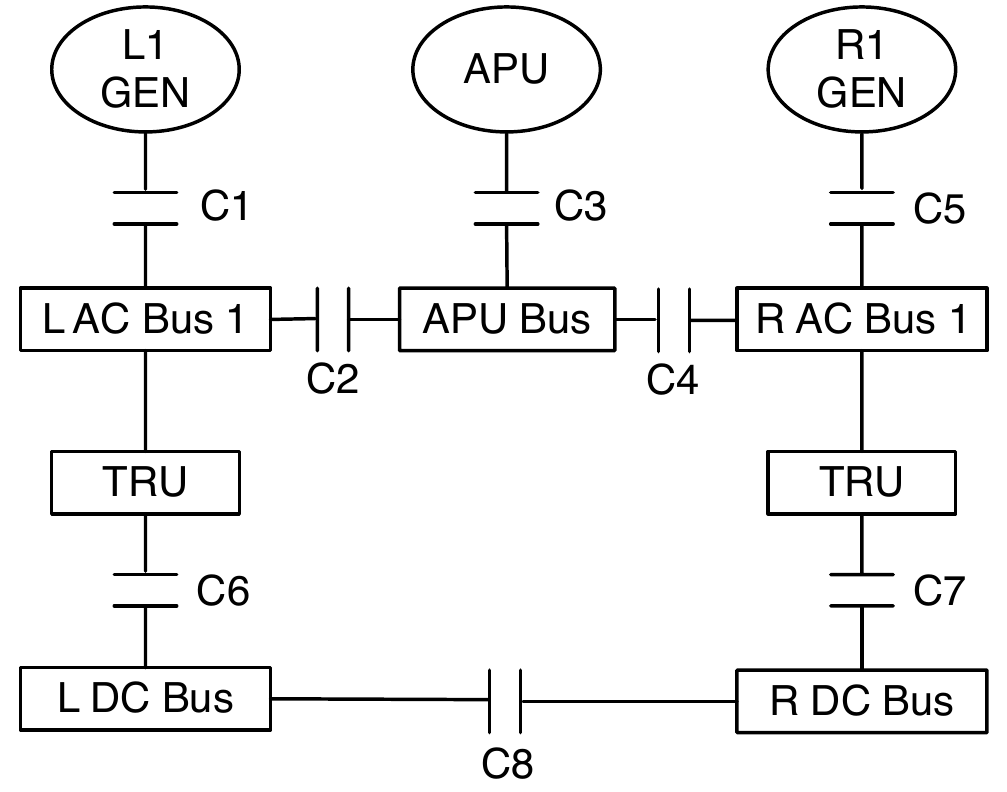}
\caption{Single-line diagram for the EPS design example.}
\label{fig:767eps}
\end{figure}

\section{The Application of the Methodology to the EPS Supervisory Control 
Design Problem}
\seclabel{application}

We illustrate our methodology on the design of the control protocol of the
primary power distribution of an EPS system~\cite{Moir08}. The primary
distribution system involves the start-up or shut-down of high-voltage
generators and APU generators as well as the configuration of contactors to
deliver power to high-voltage AC and DC buses and loads. In particular, we refer
to the topology in \figref{767eps} and the list of requirements textually
visualized in \figref{reqs}. Clearly, these requirements are just a subset of an
actual EPS specification, but they are enough to illustrate the design steps.
$R_1$ relates to system safety, $R_{2-4}$ relate to both safety and performance,
while $R_5$ is a reliability requirement.

\begin{figure}[t]
\begin{lstlisting}[mathescape]
$R_1$: No AC bus shall be simultaneously powered by more than one AC source.
$R_2$: The electric power system shall provide power with the following voltage amplitude values: 115$\pm 5$ V at 400 Hz for AC loads and 28$\pm 2$ V for DC loads.
$R_3$: L AC Bus 1 shall be powered from the first available source from the ordered list (L1 GEN, APU, R1 GEN).
$R_4$: R AC Bus 1 shall be powered from the first available source from this ordered list (R1 GEN, APU, L1 GEN).
$R_5$: The failure probability for an essential load shall be smaller than $10^{-9}$ for a 10-hour mission duration.
\end{lstlisting}
\caption{A subset of requirements for EPS design.}
\label{fig:reqs}
\end{figure}

\subsection{Capturing System Requirements}
\label{req}

We adopt mathematical formalisms to capture requirements, based on the different
domains in which the system behaviors are modeled. For example, requirements can
be expressed as automata, probabilistic constraints (e.g.~reliability
requirements), temporal logic constructs (e.g.~safety requirements),
integro-differential equations, and linear or non-linear constraints on real
numbers (e.g.~real-time performance requirements).

\begin{figure}[t]
\centering
\includegraphics[width=\columnwidth]{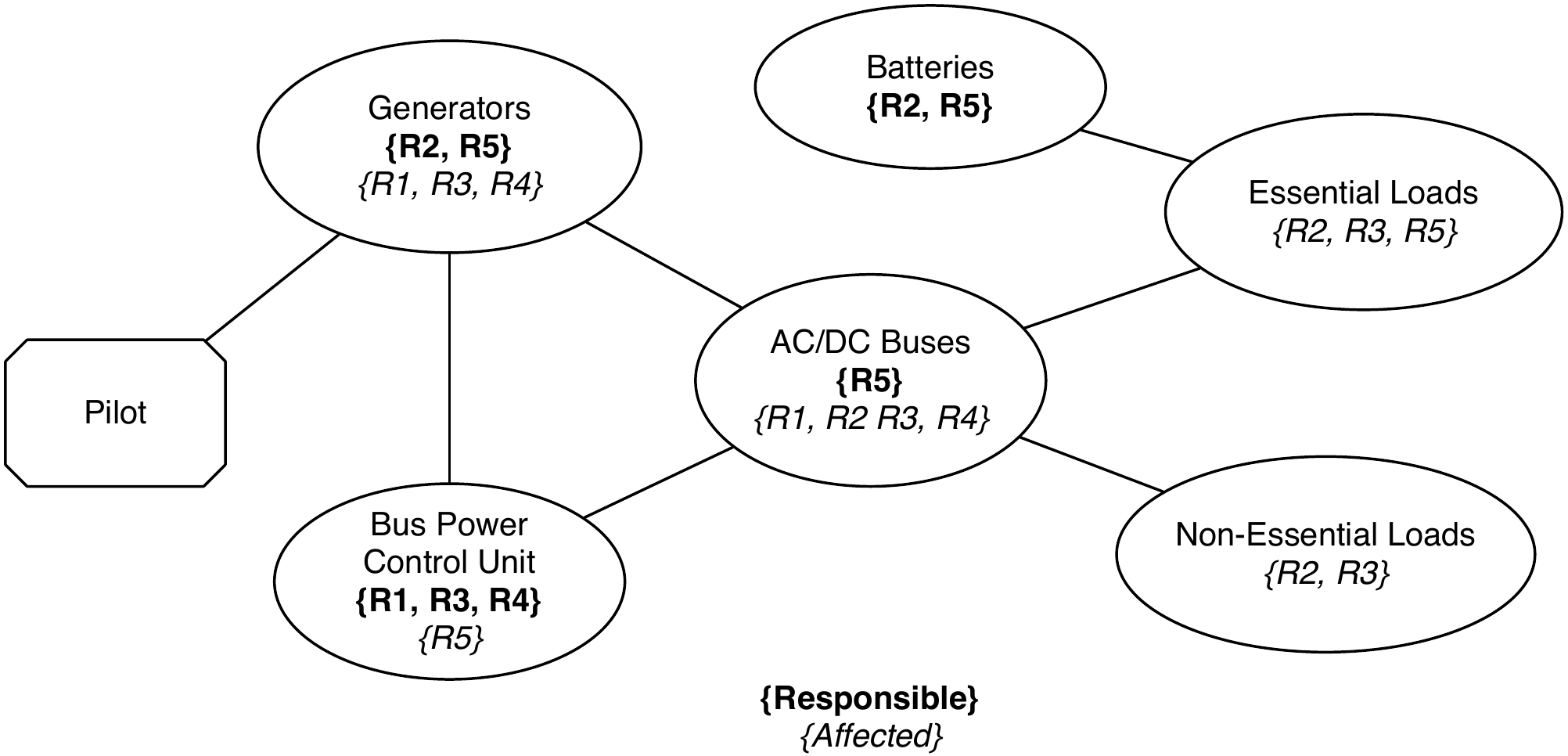}
\caption{Diagram associating system requirements with
responsible actors.}
\label{fig:usecase}
\end{figure}

To allocate requirements to components, we use the graphical representation
tools provided by SysML \emph{requirement diagrams}.
Figure~\ref{fig:usecase} shows a diagram of the system entities, highlighting
their interactions and binding each requirement with the entities that are
either responsible for its implementation or affected by it. Allocating
requirements to components allows each component to be developed independently
of the others.
For instance, the responsible entity for satisfying $R_1$ is the BPCU, while the
generators and the AC buses are the entities affected. The generators (via the
GCU) are responsible for the voltage levels specified in $R_2$, which affect
both buses and loads.  Although not shown in the figure, designers can use a few
constructs, such as \emph{derive}, \emph{refine}, \emph{verify}, \emph{trace}
and \emph{satisfy}, to establish relationships between requirements and
components so that  any modification in a requirement can be easily propagated
to all the requirements and components affected.

Based on the associations in~\figref{usecase}, the requirements for the system
components can be formalized. For example, to verify the behaviors of FSM
models, we can encode the state of a contactor $C_i$ in the topology of
\figref{767eps} with a Boolean variable, such that $C_i=1$ when the contactor is
closed. Therefore, to enforce $R_1$ on the L AC Bus 1, we can require
that no more than one path connecting $B_1$ to a power source be active at all
time.
Then, this requirement can be formalized by the following LTL property, where $\square$ is the temporal
connective ``always'' in LTL:
\begin{equation}\label{eq:ltl}
\begin{split}
\square \{\neg ((C_1 \wedge C_2 \wedge C_3) & \lor (C_1 \wedge C_2 
\wedge C_4 \wedge C_5) \\ & \lor (C_2 \wedge C_3 \wedge C_4 \wedge C_5))\}
\end{split}
\end{equation}
%
%
A similar requirement can be specified for the R AC Bus 1.

While requirement diagrams encode requirements and relations among components in
a static fashion, \emph{use case} and \emph{sequence diagrams} can be used to
further specify requirements on the component dynamics. \emph{Use case} and
\emph{sequence diagrams} are used to check system correctness, both in normal
operating conditions (e.g.~portions of mission profiles) and in the presence of
faults.
A \emph{use case diagram} describes the usage of a system by its actors (or the
environment) to achieve a specific goal. Relations between \emph{use cases} can
be expressed by using relationships such as \emph{communication},
\emph{include}, \emph{extend} and \emph{generalization}. For example, a
\emph{use case diagram} can be used to represent the interactions among the
BPCU, the power sources and the switches after a failure event. Based on our
requirements, 14 uses cases were extracted in total.
%

\begin{figure}[t]
\centering
\includegraphics[width=\columnwidth]{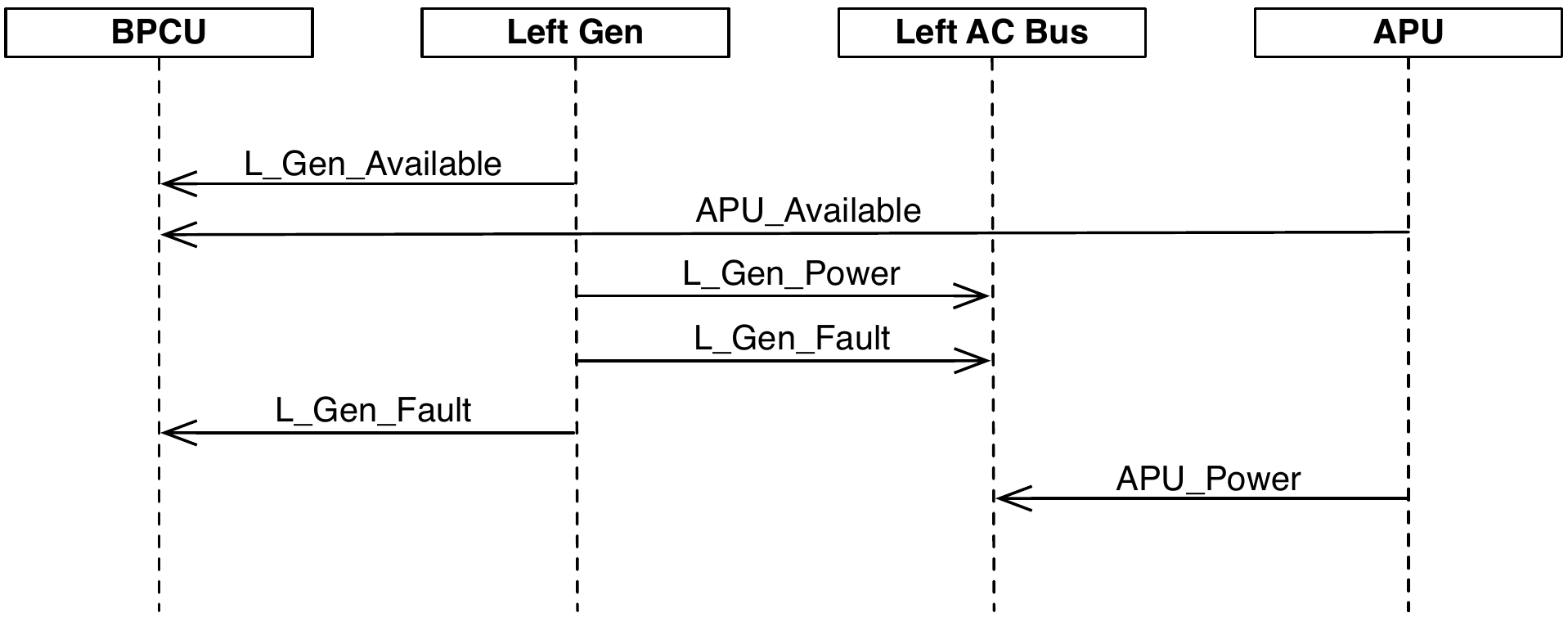}
\caption{Sequence diagram for the
verification of functional requirements for the EPS system.}
\label{fig:seq}
\end{figure}

A \emph{sequence diagram} illustrates the correct order of events in a
scenario of interest. For instance, in \figref{seq}
we specify the correct sequence of events that occurs when
the left generator in \figref{767eps} is faulty and the
BPCU is informed that the APU is available. 
\emph{Requirement, use case and sequence diagrams} can capture
system requirements without any notion of the internal
structure. 

\subsection{Hierarchical Behavioral Models}
\label{sec:behavior_mod}

\begin{figure}[t]
\centering
\includegraphics[width=\columnwidth]{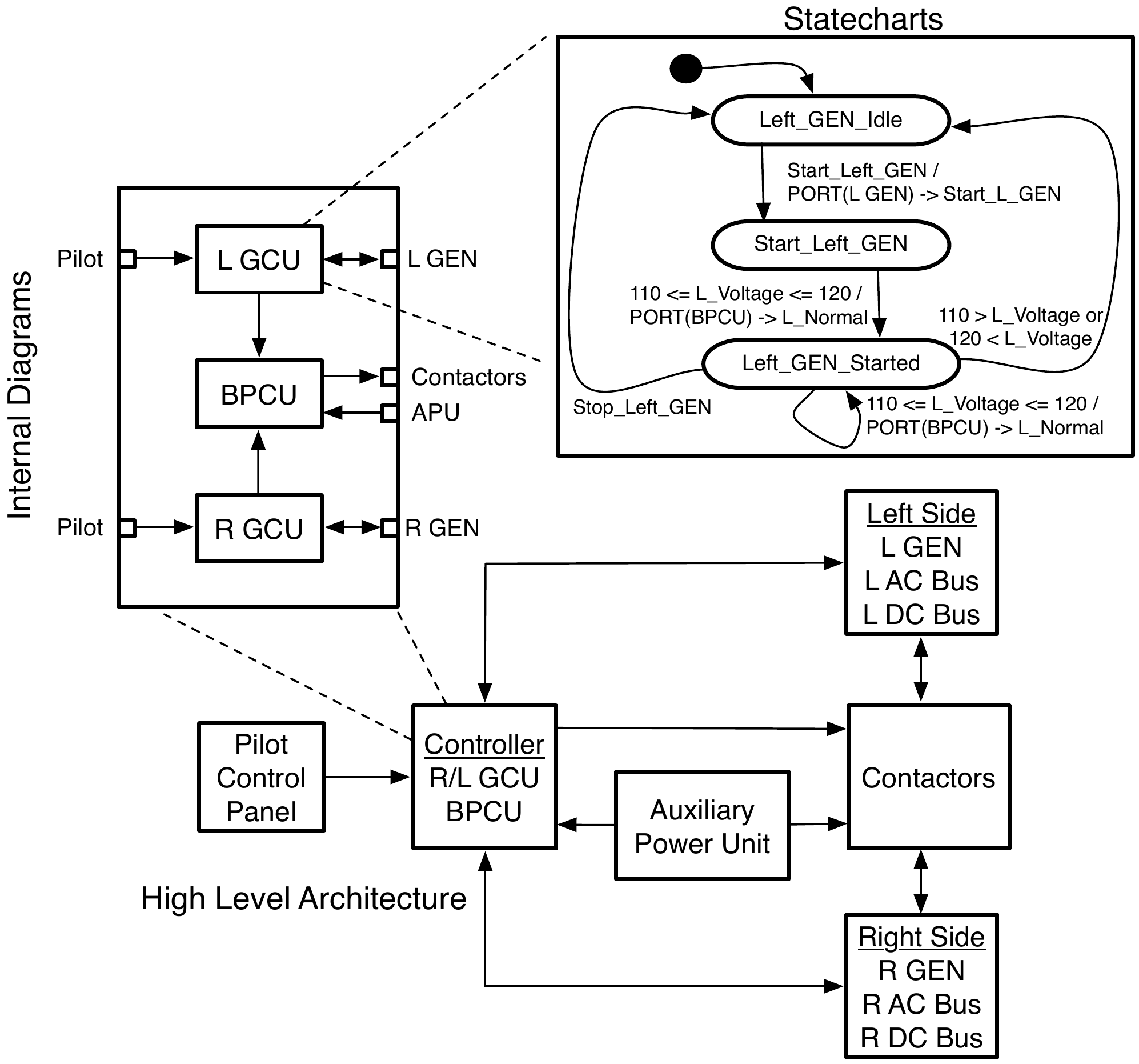}
\caption{EPS structure and state-machine diagram.}
\label{fig:architecture}
\end{figure}

To execute requirements at a high level of abstraction, finite-state machines
are implemented using SysML \emph{structure diagrams}, \emph{state-machine
diagrams} or \emph{Matlab Stateflow diagrams}.
A SysML \emph{structure diagram} consists of a set of blocks that are connected
together using ports and can be organized hierarchically. The behavior of each
block can  be further described by a finite-state transition system. For
example, the overall EPS architecture is visualized in \figref{architecture},
showing the FSM model of a generator and its GCU as an inset. The model consists
of 6 high-level blocks and 11 internal state-machine diagrams, corresponding to
approximately 1 billion states for the overall (flattened) system.

\begin{figure}[t]
\centering
\includegraphics[width=\columnwidth]{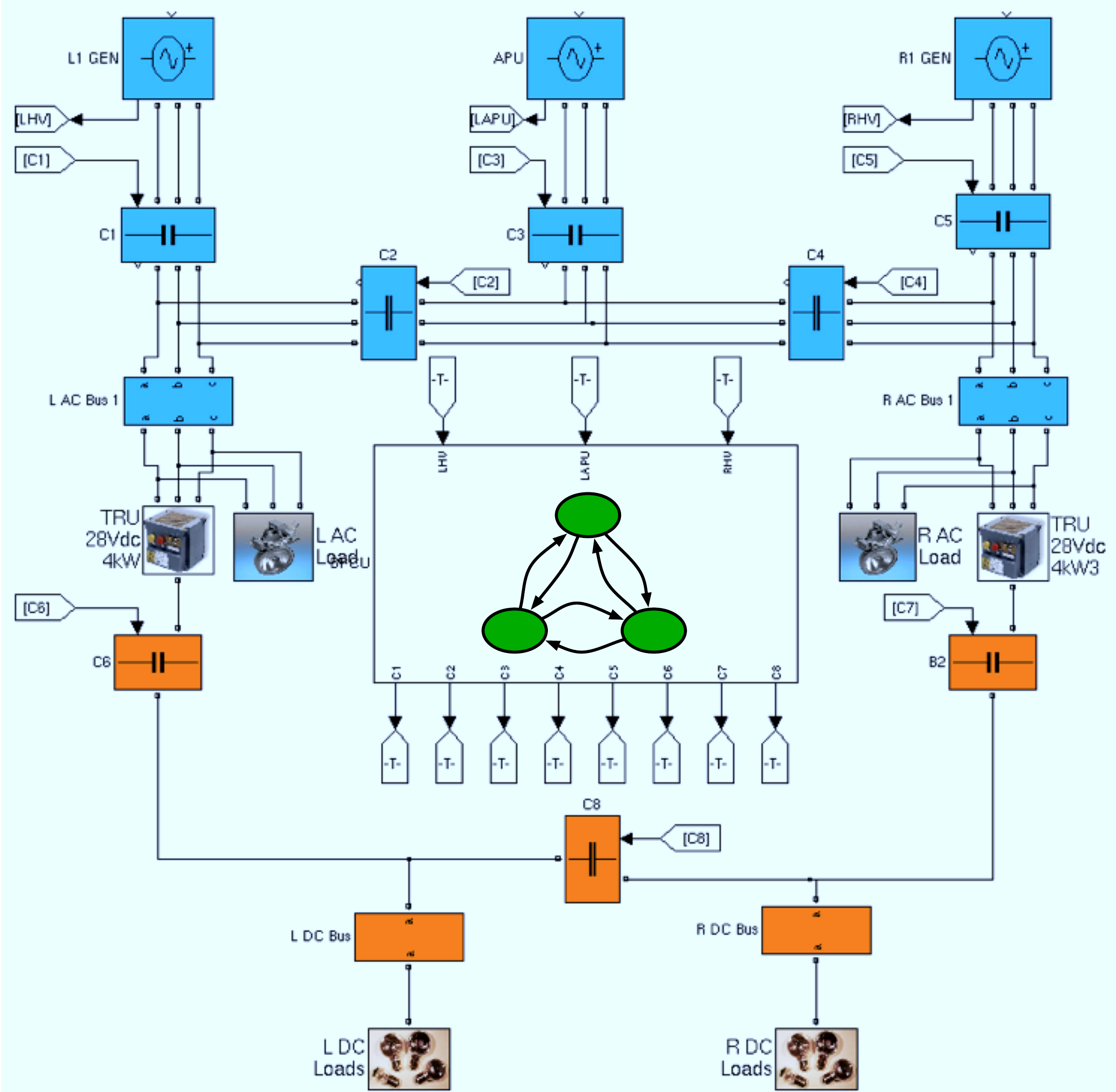}
\caption{EPS Simulink model.}
\label{fig:simulinkEPS}
\end{figure}

Continuous-time models are implemented in Simulink, by exploiting the
SimPowerSystems extension.
As an example, the continuous-time model for the
generator consists of a mechanical engine (turbine), a three-phase synchronous
generator, and the GCU, driving the field voltage of the generator, thus
refining the three-state model in \figref{architecture}.
In addition to timing properties, CT models allow measuring current and voltage levels
at the different circuit loads, as specified, for instance, by requirement
$R_2$. 
Figure~\ref{fig:simulinkEPS} shows the overall Simulink hybrid model, which refines
the SysML structure diagram described above. In this model, CT abstractions
of the generators, contactors, loads and power converters interface with a Stateflow
implementation of the BPCU. The BPCU state-machine consists of 10 states and
41 transitions. 

\subsection{Design Example} \seclabel{results}

\begin{figure}[t]
\centering
\includegraphics[width=\columnwidth]{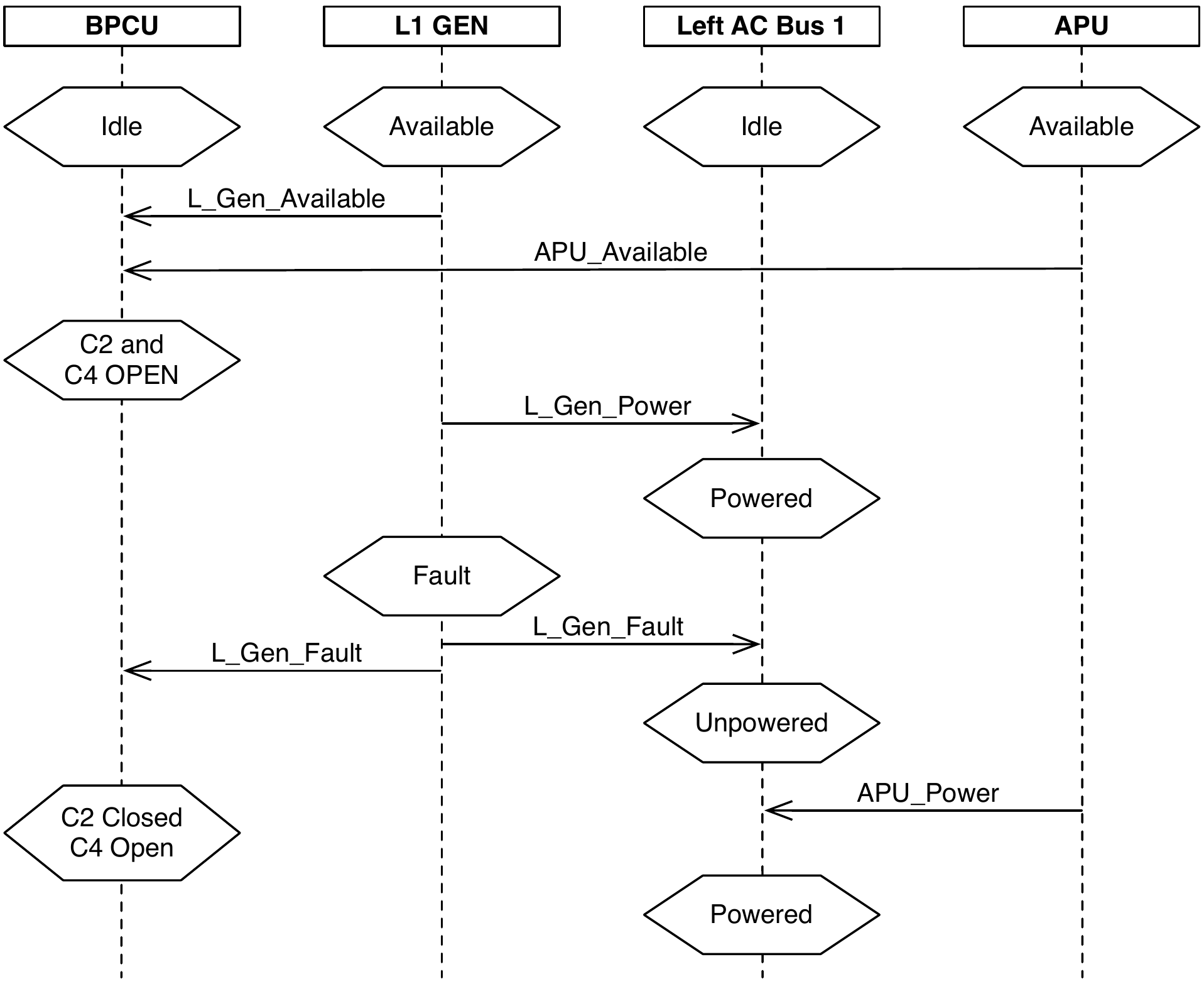}
\caption{Simulated sequence diagram.}
\label{fig:seq_gen}
\end{figure}

For the example in this paper, we manually designed a controller state machine
for the topology in \figref{767eps}. A large set of requirements, such as safety
and priority constraints, can be efficiently captured via finite-transition
systems, executed using an event-driven simulator, and analyzed using a model
checker. For instance, properties as the one in~\eqref{eq:ltl} can be verified
against our controller model in less than one second using NuSMV~\cite{nusmv}. 
In this example, we also checked such requirements on the SysML model in
\figref{architecture}, by using IBM Rational Rhapsody~\cite{ibm}
to implement, simulate and analyze the models. Model compilation took
approximately 1 minute on a 2.53-GHz Intel Core i5 processor with 4 GB of
memory. For instance, \figref{seq_gen} shows the sequence diagram generated by
simulating our initial design for the same scenario as in \figref{seq},
highlighting the evolution of the different state machines. We could then
automatically compare the sequence diagram in \figref{seq} with the one in
\figref{seq_gen} within Rhapsody, and show that the simulated diagram is indeed
correct with respect to the specification. Alternatively, in case a violation is
detected, the simulated traces can be used as counterexamples to further refine
the design.

Once the safety requirements were verified, we needed to assess the real-time
performance of candidate BPCU designs. For this purpose, we simulated a
Stateflow version of the designed controller together with the continuous-time
model of the power generation and distribution network. Since buses were now
modeled as transmission lines connected to electrical loads, we could capture
the transient behaviors of currents and voltages. Moreover, we could also
introduce parameters related to the hardware implementation of the control
algorithm, such as clock frequency, under the assumption of a centralized
synchronous controller. Our requirements were also refined to be consistent with
the signals of the hybrid model. At this level, we verified our properties by
implementing observers, capable of monitoring voltage and current waveforms.
A 2-second simulation with a step size of 10$\mu$s took approximately 60
seconds. As an example of design refinement, we consider a scenario in which L1 GEN in
\figref{767eps} is powering both the L and R AC buses, and the APU generator becomes
available and needs to be routed to power R AC Bus 1 as required by $R_4$.
Then, to satisfy requirement $R_1$ on the APU Bus in \figref{767eps}, we require
$C_3$ to be closed only after $C_2$ is open. We then consider that $C_2$ is
actually ``open'' only when the current through it decays below a safety
threshold, approximately equal to $10\%$ of its original value. As shown at the
top of \figref{simulink}, our initial BPCU design did not satisfy $R_1$, since
$C_3$ was ``on'' before $C_2$ could be considered safely ``off''. Therefore, we
had to refine the original design to explicitly monitor the current through
$C_2$ and guarantee a deterministic delay before $C_3$ is closed. Simulated
waveforms from our final design are shown at the bottom of \figref{simulink}.

\section{Discussion and Conclusions} \seclabel{conclusions}

\begin{figure}[t]
\centering
\includegraphics[width=\columnwidth]{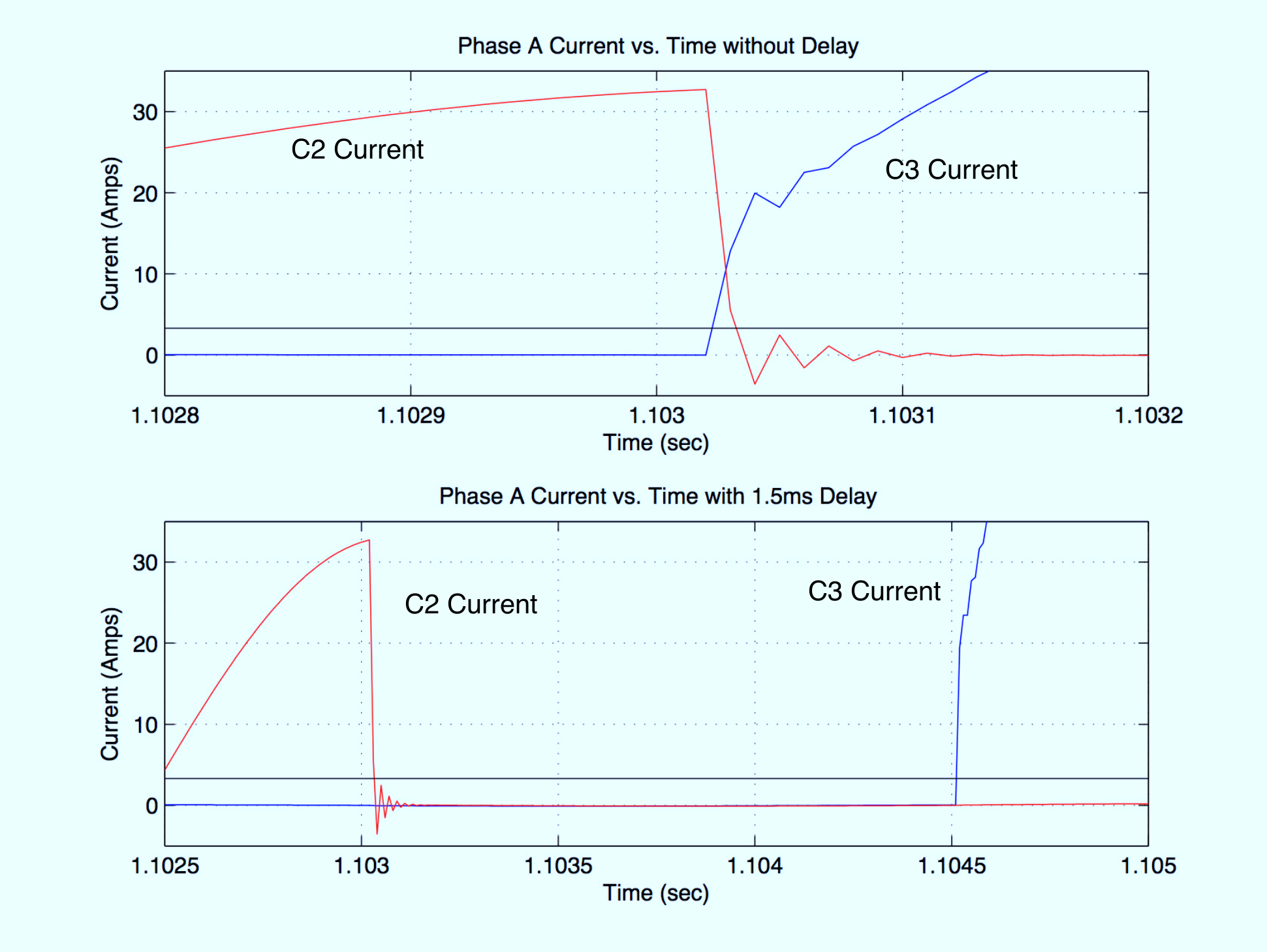}
\caption{Transient behavior of the currents through
$C_2$ and $C_3$ in \figref{767eps} c) for the initial
BPCU design (top) and the final one (bottom).}
\figlabel{simulink}
\end{figure}

We have introduced a platform-based design methodology for virtual prototyping
of supervisory control protocols in aircraft electrical power systems. We have
detailed the modeling infrastructure that supports our methodology and
demonstrated some of the steps on a prototype design. We have used SysML
requirement diagrams to capture the top-level system specifications,
state-machine diagrams to enable efficient functional verification of the
control protocol, and hybrid models in Simulink, to verify the satisfaction of
real-time performance constraints. 

As a future work, we plan to develop an environment for automatic deployment of
the proposed methodology, including the definition of performance metrics to
validate it as well as the implementation of concepts from contract-based
design~\cite{Sangiovanni-Vincentelli2012} to address the challenges of
coordinating diverse formalisms and checking consistency across diverse
models maintained in multiple tools.

\section{Acknowledgments}
The authors wish to acknowledge Rich Poisson from United Technologies
Corporation (UTC) for helpful discussions.
This work was supported in part by IBM and UTC via the iCyPhy consortium and by
the TerraSwarm Research Center, one of six centers of STARnet, a Semiconductor
Research Corporation program sponsored by MARCO and DARPA.

\bibliographystyle{IEEEtran}

\begin{thebibliography}{}
\providecommand{\url}[1]{#1}
\csname url@samestyle\endcsname
\providecommand{\newblock}{\relax}
\providecommand{\bibinfo}[2]{#2}
\providecommand{\BIBentrySTDinterwordspacing}{\spaceskip=0pt\relax}
\providecommand{\BIBentryALTinterwordstretchfactor}{4}
\providecommand{\BIBentryALTinterwordspacing}{\spaceskip=\fontdimen2\font plus
\BIBentryALTinterwordstretchfactor\fontdimen3\font minus
  \fontdimen4\font\relax}
\providecommand{\BIBforeignlanguage}[2]{{%
\expandafter\ifx\csname l@#1\endcsname\relax
\typeout{** WARNING: IEEEtran.bst: No hyphenation pattern has been}%
\typeout{** loaded for the language `#1'. Using the pattern for}%
\typeout{** the default language instead.}%
\else
\language=\csname l@#1\endcsname
\fi
#2}}
\providecommand{\BIBdecl}{\relax}
\BIBdecl

\end{thebibliography}


\begin{thebibliography}{10}
\providecommand{\url}[1]{#1}
\csname url@samestyle\endcsname
\providecommand{\newblock}{\relax}
\providecommand{\bibinfo}[2]{#2}
\providecommand{\BIBentrySTDinterwordspacing}{\spaceskip=0pt\relax}
\providecommand{\BIBentryALTinterwordstretchfactor}{4}
\providecommand{\BIBentryALTinterwordspacing}{\spaceskip=\fontdimen2\font plus
\BIBentryALTinterwordstretchfactor\fontdimen3\font minus
  \fontdimen4\font\relax}
\providecommand{\BIBforeignlanguage}[2]{{%
\expandafter\ifx\csname l@#1\endcsname\relax
\typeout{** WARNING: IEEEtran.bst: No hyphenation pattern has been}%
\typeout{** loaded for the language `#1'. Using the pattern for}%
\typeout{** the default language instead.}%
\else
\language=\csname l@#1\endcsname
\fi
#2}}
\providecommand{\BIBdecl}{\relax}
\BIBdecl

\bibitem{Moir08}
I.~Moir and A.~Seabridge, \emph{Aircraft Systems: Mechanical, Electrical and
  Avionics Subsystems Integration. Third Edition}.\hskip 1em plus 0.5em minus
  0.4em\relax Chichester, England: John Wiley and Sons, Ltd, 2008.

\bibitem{Pinto2010}
A.~Pinto, S.~Becz, and H.~M. Reeve, ``Correct-by-construction design of
  aircraft electric power systems,'' in \emph{AIAA Aviation Technology,
  Integration, and Operations Conf.}, 2010.

\bibitem{ASV:quovadis}
A.~Sangiovanni-Vincentelli, ``Quo vadis, {SLD}? {R}easoning about the trends
  and challenges of system level design,'' \emph{Proc. IEEE}, no.~3, pp.
  467--506, 2007.

\bibitem{nuzzo:kluPipeline}
P.~Nuzzo, F.~De~Bernardinis, and A.~Sangiovanni~Vincentelli, ``Platform-based
  mixed signal design: {O}ptimizing a high-performance pipelined {ADC},''
  \emph{Analog Integr. Circuits Signal Process.}, vol.~49, no.~3, pp. 343--358,
  2006.

\bibitem{sysml}
\BIBentryALTinterwordspacing
\emph{OMG Systems Modeling Language}. [Online]. Available:
  \url{http://www.sysml.org/}
\BIBentrySTDinterwordspacing

\bibitem{Krus00}
P.~Krus and J.~Nyman, ``Complete aircraft system simulation for aircraft design
  - paradigms for modelling of complex systems,'' in \emph{Int. Congress of
  Aeronautical Sciences}, 2000.

\bibitem{Bals05}
J.~Bals, G.~Hofer, A.~Pfeiffer, and C.~Schallert, ``Virtual iron bird - a
  multidisciplinary modelling and simulation platform for new aircraft system
  architectures,'' in \emph{German Aerospace Conference}, 2005.

\bibitem{MOET09}
\BIBentryALTinterwordspacing
T.~Jomier \emph{et~al.}, ``Final {MOET} technical report,'' Tech. Rep., Dec.
  2009. [Online]. Available: \url{http://www.eurtd.com/moet/}
\BIBentrySTDinterwordspacing

\bibitem{modelica}
\BIBentryALTinterwordspacing
\emph{Modelica Language}. [Online]. Available: \url{http://www.modelica.org}
\BIBentrySTDinterwordspacing

\bibitem{huynh2006}
T.~V. Huynh and J.~S. Osmundson, ``A systems engineering methodology for
  analyzing systems of systems using the {S}ystems {M}odeling {L}anguage
  ({SysML}),'' in \emph{{S}ystem of {S}ystems Engineering Conf.}, 2006.

\bibitem{Guo2009}
G.~Yue and R.~P. Jones, ``A study of approaches for model based development of
  an automotive driver information system,'' in \emph{IEEE Systems Conf.}, Mar.
  2009, pp. 267--272.

\bibitem{Johnson2007}
T.~A. Johnson, C.~J.~J. Paredis, R.~Burkhart, and J.~M. Jobe, ``Modeling
  continuous system dynamics in {S}ys{ML},'' in \emph{ASME International
  Mechanical Engineering Congress and Exposition}, Nov. 2007.

\bibitem{Michalko2008}
R.~G. Michalko, ``Electrical starting, generation, conversion and distribution
  system architecture for a more electric vehicle,'' \emph{US Patent 7,439,634
  B2}, Oct. 2008.

\bibitem{arp}
{The Aerospace Recommended Practice}, ``{ARP4754: Guidelines for Development of
  Civil Aircraft and Systems},'' 2012.

\bibitem{Wongpiromsarn11}
T.~Wongpiromsarn, U.~Topcu, N.~Ozay, H.~Xu, and R.~M. Murray, ``{T}u{L}i{P}: a
  software toolbox for receding horizon temporal logic planning,'' in
  \emph{Intl. Conf. Hybrid Systems: Computation and Control}, 2011.

\bibitem{CDC_Mumu}
H.~Xu, U.~Topcu, and R.~M. Murray, ``A case study on reactive protocols for
  aircraft electric power distribution,'' in \emph{Int. Conf. Decision and
  Control}, 2012.

\bibitem{Pnueli77}
A.~Pnueli, ``The temporal logic of programs,'' in \emph{Proc. Symp. on
  Foundations of Computer Science}, Nov. 1977, pp. 46--57.

\bibitem{Hang11}
C.~Hang, P.~Manolios, and V.~Papavasileiou, ``Synthesizing cyber-physical
  architectural models with real-time constraints,'' in \emph{Proc. Int. Conf.
  Comput.-Aided Verification}, Dec. 2011.

\bibitem{nusmv}
A.~Cimatti, E.~Clarke, E.~Giunchiglia, F.~Giunchiglia, M.~Pistore, M.~Roveri,
  R.~Sebastiani, and A.~Tacchella, ``{NuSMV Version 2: An OpenSource Tool for
  Symbolic Model Checking},'' in \emph{Proc. Int. Conf. Computer-Aided
  Verification}, 2002.

\bibitem{ibm}
\BIBentryALTinterwordspacing
\emph{IBM Rational Rhapsody}. [Online]. Available:
  \url{http://www-01.ibm.com/software/rational/products/rhapsody/developer/}
\BIBentrySTDinterwordspacing

\bibitem{Sangiovanni-Vincentelli2012}
A.~Sangiovanni-Vincentelli, W.~Damm, and R.~Passerone, ``{Taming Dr.
  Frankenstein: Contract-Based Design for Cyber-Physical Systems},''
  \emph{European Journal of Control}, vol.~18, no.~3, pp. 217--238, Jun. 2012.

\end{thebibliography}

\begin{IEEEbiography}
[{\includegraphics[width=1in,height=1.25in,clip,keepaspectratio]{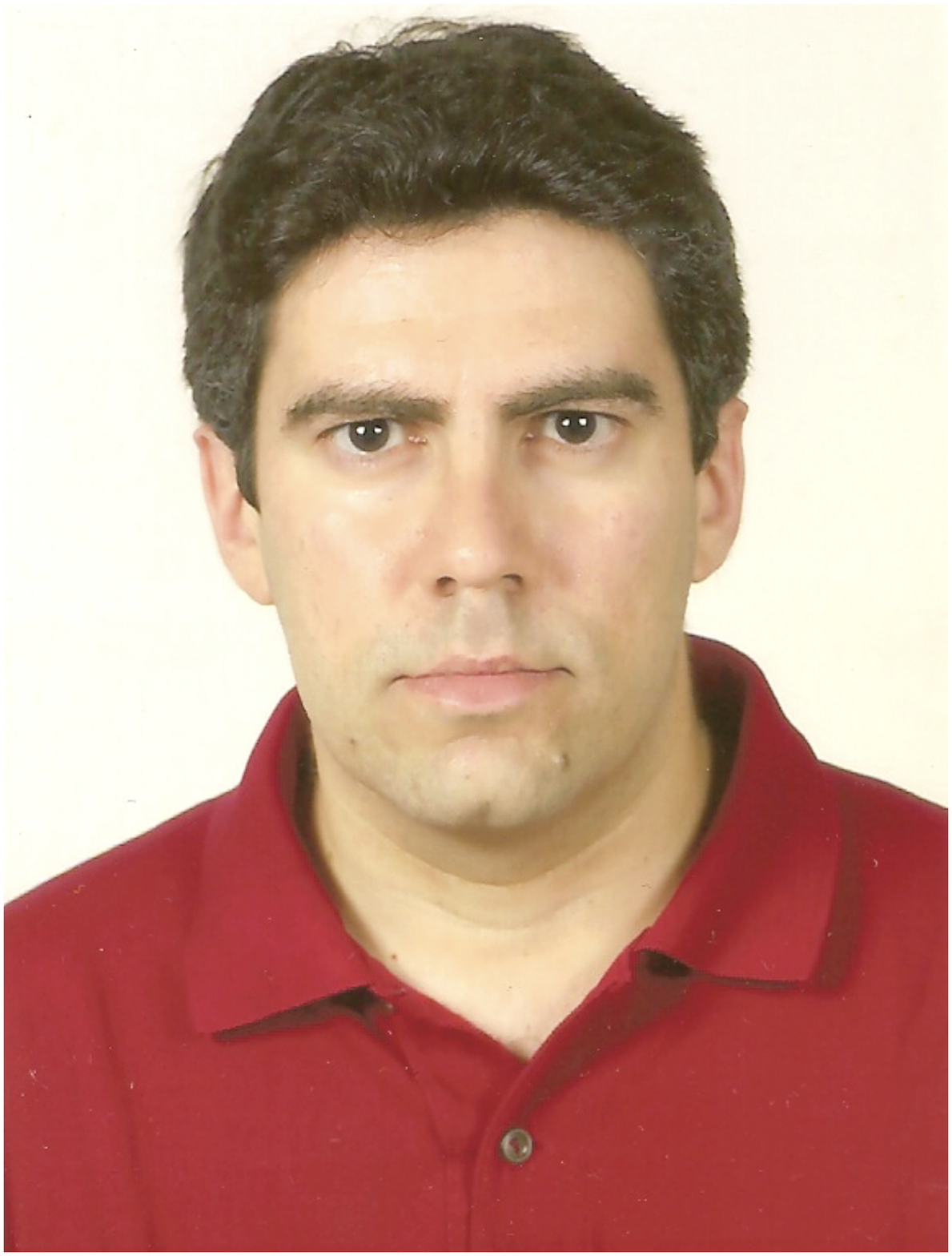}}]{
Pierluigi Nuzzo \textnormal{received the M.Sc. degree in electrical engineering from the
University of Pisa, Italy, in 2003. He is currently with the University of
California at Berkeley, where he is working toward the Ph.D. degree in
electrical engineering and computer sciences. His research interests include
methodologies and tools for the design of cyber-physical systems and
mixed-signal integrated circuits.
He held research positions at the University of Pisa and IMEC, Leuven, Belgium,
working on the design of energy-efficient A/D converters and frequency
synthesizers for reconfigurable radio. He received the First Place in the 2006
DAC/ISSCC Design Competition, the U.C. Berkeley EECS fellowship in 2008 and the
IBM Ph.D. Fellowship in 2012.}
}
\end{IEEEbiography}

\begin{IEEEbiography}
[{\includegraphics[width=1in,height=1.25in,clip,keepaspectratio]{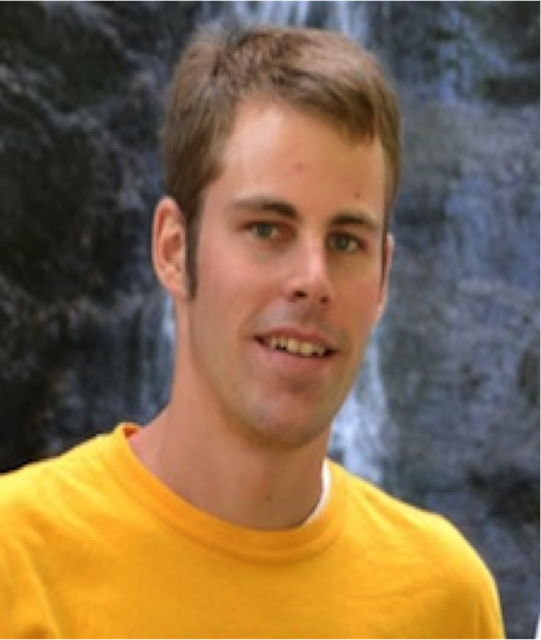}}]{
John B. Finn \textnormal{is a Ph.D. student at the University of California,
Berkeley, advised by Professor Alberto Sangiovanni-Vincentelli. In 2010, he
received his bachelor's degree (BSEE) with highest distinction in electrical
engineering from Purdue University, West Lafayette, IN. His research interests
include the design, optimization and verification of cyber-physical systems as
well as system modeling and analysis.  He has accumulated three years work
experience with internships at United Technologies Aerospace Systems in
Rockford, IL, BAE Systems in Nashua, NH, United Technologies Research Center in
East Hartford, CT and The Advanced Laboratory on Embedded Systems in Rome,
Italy.}
}
\end{IEEEbiography}

\begin{IEEEbiography}
[{\includegraphics[width=1in,height=1.25in,clip,keepaspectratio]{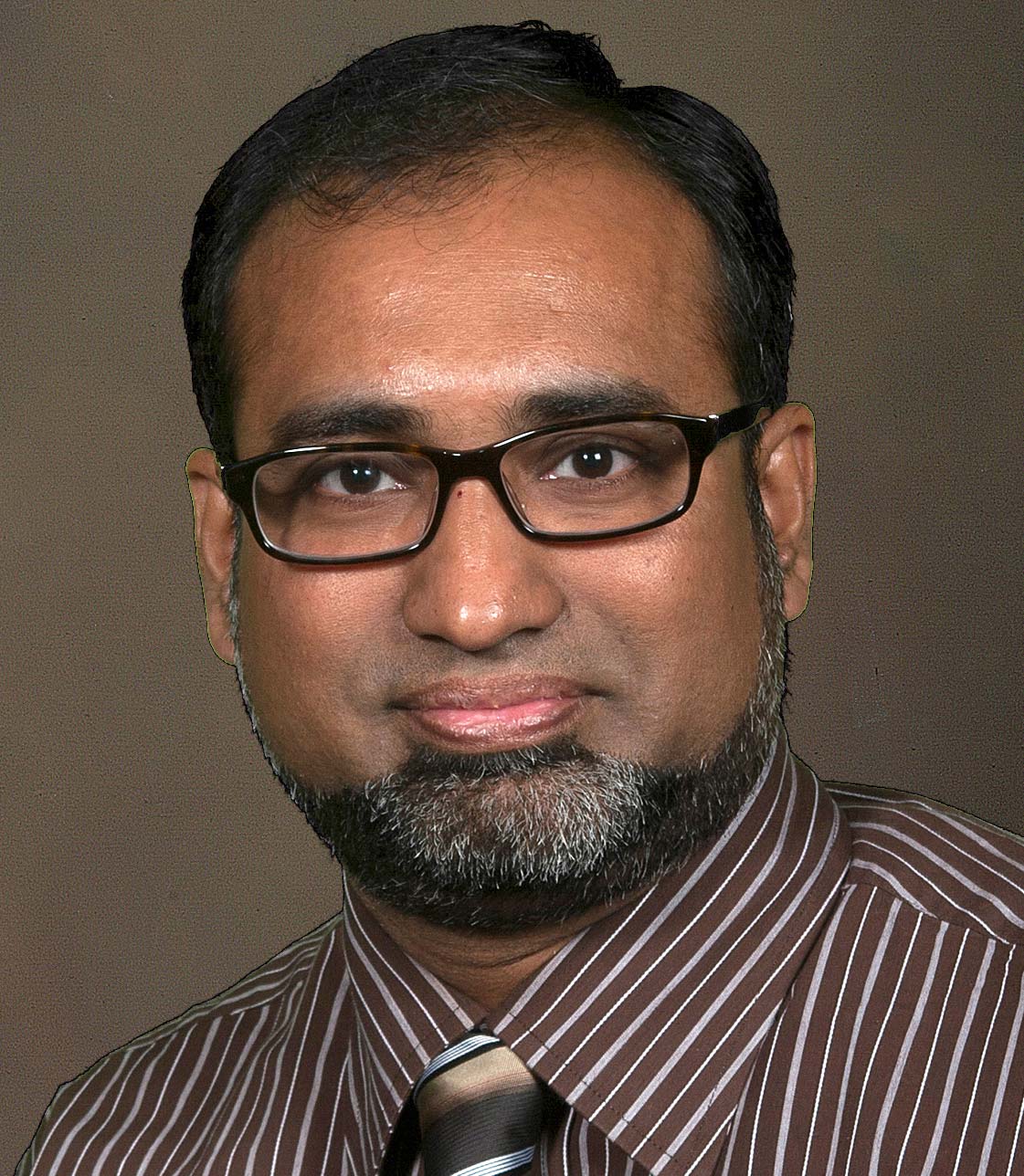}}]{
Mohammad Mostafizur Rahman Mozumdar} is a tenure track faculty in the Electrical
Engineering department of the California State University at Long Beach and an
ex-postdoc from the University of California, Berkeley. He received a Ph.D. in
electronics and communication engineering from Politecnico di Torino, Italy. His
ideas on model-based design for sensor networks made profound impact on
engineering and industrial communities and have been published in book chapters,
renowned journals, reputed conference proceedings and major scientific
magazines. Dr. Mozumdar's research interests include methodologies and tools for
sensor network design, energy-efficient building information and control system
design, cloud computing, methodologies for the design of distributed embedded
systems and cyber-physical systems subject to real-time, safety and reliability
constraints.
\end{IEEEbiography}

\begin{IEEEbiography}
[{\includegraphics[width=1in,height=1.25in,clip,keepaspectratio]{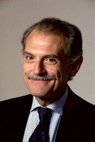}}]
{Alberto Sangiovanni Vincentelli} received the Laurea degree in electrical
engineering and computer sciences from the Politecnico di Milano, Italy in 1971.
He currently holds the Edgar L. and Harold H. Buttner Chair of Electrical
Engineering and Computer Sciences at the University of California at Berkeley.
He was a co-founder of Cadence and Synopsys, he is the Chief Technology Adviser
of Cadence, a member of the Board of Directors of Cadence and a member of the
Science and Technology Advisory Board of General Motors. He is an author of over
880 papers and 15 books in the area of design tools and methodologies,
large-scale systems, embedded systems, hybrid systems and innovation. He has won
numerous awards, including, among others, the IEEE/RSE Wolfson James Clerk
Maxwell Award, the Kaufman Award and the ACM/IEEE Richard Newton Technical
Impact Award in Electronic Design Automation.
\end{IEEEbiography}

\end{document}